# The differences between natural and artificial life
## *Towards a definition of life*

**Andrzej Gecow**
Centre for Ecological Research Polish Academy of Sciences
M. Konopnickiej 1, Dziekanow Lesny, 05-092 Lomianki, Poland
Bogusławskiego 4/76, 01-923 Warsaw, Poland
gecow@op.pl

**Abstract** It is high time to openly and without finalism define the dangerous but needed term 'purposeful information', whose quantity is an Eigen information value. Using the term 'biological information' in its instead forces one into an uncomfortable detour. I propose such a definition based on the generalized notions of 'information' and 'encoding'. Next, the properties of the spontaneous process of collecting purposeful information are investigated. In effect, the properties of this process: the goal 'continuation of existence', reproduction and Darwinian mechanism are derived which suggest, that it is the natural life process. A 'natural identity criterion' appears in this process for an evolving object, that is connected to a 'small change tendency'. Likewise, 'hereditary information' is defined. Artificial life is constructed by living objects, is a part of natural life process and its properties are not an effect of its internal restrictions but of external assumptions.

## 1 Introduction

### 1.1 Difference between Natural and Artificial Life and the Philosophic Aim of the Article

Natural and artificial life differ, by definition, in their origin and this implies the nature of the remaining differences between them. The aspect of the source of coming into existence seems to be irrelevant to the main properties of a created thing. However, it occurs that the main characteristics of the process of natural life can be derived from an assumption that the process is spontaneous.

Lynne Baker (2008) defines: 'Artifacts are objects intentionally made to serve a given purpose; natural objects come into being without human intervention'. Philosophers debate on the ontological differences between artifacts (or other mind-dependent objects) and natural objects, but their point of view is the human point of view: *i.e.* that of the designer, user and researcher of artifacts. They use the mind which is simultaneously the observed and the observer, and this may lead to a number of problems. She rightly remarks later on: 'Of course the existence of artifacts depends on us: but we are part of nature'. If we look as non-human observer, then her artifacts are parts of human civilization, which itself is a part of life evolution. Their status is thereby the same as that of a bird nest, ant-hill, honeycomb, or even a snail shell. However, we can travel further along this pathway to include a tooth, a bone, the heart (why not?), and next, a flagellum in a cell or even a protein or tRNA and so on. Along the way, there is no threshold which can be a base of the artificiality criterion. The mind is not a good threshold. Moreover, this way exhausts all the carriers of life. What, or who, in such a case, holds the purposes and creates artifacts? Dawkin's selfish gene makes proteins as its artificial tools. Should genes at the end of the pathway be considered natural? We can adopt such a convention, however, it is not nearly as adequate as hypercycles, all of whose parts are equally important. It means that in extreme cases, a gene may be the tool of the protein as well (this in spite of genetic dogma).

The purposefulness needed in defining artifacts is an objective effect of Darwinian mechanism, and all that results from this mechanism is similarly and objectively artificial. Any part of a living object is an artifact fulfilling a broader interpretation of Baker's definition of artifacts mentioned above, where 'intentionally' and 'given purpose' are human-related names of real elements of Darwinian mechanism, *i.e.* natural selection. By its definition, 'selection' is **intentional**, but in the first simple approximation, natural selection is an **automatic** elimination of objects not fit for existence (*i.e.* - for the 'purpose') from a set of ones of random (natural, spontaneous) origin.

What is the result of the assumption that natural life can be not artificial? This assumption means that natural life can be spontaneous, *i.e.* not 'intentionally made to serve a given purpose', and this way purposefulness does not enter the life. However, we await that the purpose and purposeful information lie in the base of the life. Such awaiting results from the above analysis of the objective range of artifacts (which contain a purpose) and from the old remark that purposefulness is truly seen in the range of life. Living objects seem to contain lot of collected purposeful information. Summarizing, is this possible that purposeful information is collected spontaneously? If it is possible, then how this process looks like? It will occur that this process is very similar to natural life.

### 1.2  Purposefulness and Purposeful (Biological) Information in the Range of Life

The detection purposefulness anywhere around us is an extrapolation of human behaviour and social interaction out of this area. As such it was removed from the scientific understanding of nature. But within the structural, functional and behavioural range of living objects an explanation based on purpose turns out to be so natural and effective that it cannot be removed. If, in an area of human activity, a reference to the assumed purpose is natural, expected and creates no objections, a biological explanation making use of purposefulness needs to point out to the mechanism that creates such a purpose. Therefore, all that is artificial, including artificial life, can (and must, see ch.1.1) be purposeful but if we assume that the purposefulness of natural life has the same causes as artificial life, then we must agree that the source of it can be only God. Science should ascertain whether other solutions exist. Darwin discovered a source of purposefulness: random (without direction to purpose, *i.e.* spontaneous) variability (origin of new objects) and natural selection, *i.e.* the automatic elimination of new objects which do not effectively serve a given purpose (existence). Later theories like 'Modern evolutionary synthesis' do not add anything to this aspect. Colin Pittendright (1958) introduced the notion 'teleonomy' for purposefulness in the range of life to distinguish it from mystic teleology.

In the term '*information*', particularly '*biological information*', one intuitively detects an aspect of purposefulness, but Shannon theory lacks them. Eigen (1971) tried to grasp this aspect by introducing, additionally to '*information quantity*' for aspect present in Shannon theory, an '*information value*' for aspect of usefulness for a receiver connected to the receiver's purposes, but this promising '*value*' has never been well defined. Millikan (1984) summarized the purposefulness aspect in her teleosemantic theory. A student of Eigen, Küppers (1986), analyzing the 'fundamental notion – biological information' introduces its three dimensions: syntactic (Shannon's), pragmatic and semantic (the latter two being teleosemantic). Typically, in the definition of information, the flow of information from a source to a receiver is considered. Shea (2007) following Millikan (1984) emphasizes the roles of both producers and consumers. Jablonka (2002) currently proposes what I believe to be the most advanced definition of biological information. She bases on consumer mechanisms alone, *i.e.* on the causal effect of the process for the receiver. Her definition concerns the semantic aspect: 'a source becomes an informational input when an interpreting receiver can react to the form of the source (and variations in this form) in a functional manner'. The word 'can' helps to avoid an a'posteriori structure. Despite Jablonka's efforts, the purposefulness aspect remains. However, her definition can be used more broadly than Jablonka suggests, out of the biological and technical spheres, *e.g.* in physics to describe phenomena which have nothing to do with purpose.

My approach is similar but I start from the broader notion of 'information' without purposefulness, and I detach the dynamic aspect of information flow to the notion of 'coding'. The notion of `code' (or `coding') in technology is used for a small, specific area which contains a coding mechanism. Here, it becomes crucial and autonomous. A 'natural code' is defined using these two expanded notions for a description of an isolated physical system. Analysis of the decoding of information encoded by the natural code defines the notion of `purpose' and next `purposeful information'. Definitions obtained in this way correctly and precisely describe the popular meaning of these notions.

The lack of any current and serious analysis of the notion of `purpose', especially in the context of `purposeful information', forces one to go on a uncomfortable detour. It is high time to propose an exact theory, which correctly defines and interprets its own terms, including Eigen's '*information value*', *purpose* and *purposeful information* without committing the error of finalism. I hope that I have achieved this as one coherent entity with the abovementioned source of purposefulness, the Darwinian mechanism. Earlier, I announced this view in (Gecow 2008b). I can repeat following Chaitin (1979) 'This paper is intended as a contribution to von Neumann's program of formulating mathematically the fundamental concepts of biology in a very general setting, *i.e.* in highly simplified model universes.'

In this paper I derive the Darwinian mechanism and reproduction starting from the spontaneous origin of purposefulness in the form of purposeful information. The elimination test displays a tendency to prefer very small changes from which the natural identity criterion emerges. This criterion creates an evolving object (see Gecow 2009b). Reproduction appears as a needed property of the long spontaneous process of collecting purposeful information and as the first purposeful information. Different types of purposeful information need different mechanisms to information growth. When forms of activity of purposeful information are analyzed in more detail, hereditary information is defined. Interpretation of this theoretical notion is broader end more precise even than that of Jablonka & Lamb (2005).

'Biological information' as a general and informal term without a precise definition has evolved and commonly becomes a synonym of genetic information. It is a strong simplification resulting from a 'first approximation' stage of inheritance theory which limits itself to genetic inheritance only. Here, also Jablonka and Lamb (1989, 1995, 2005) are leaders of a 'second approximation' which remarks on other channels of hereditary information. Jablonka (2002) wrote: `Clearly, according to my definition, genes have no theoretically privileged informational status. They are just one type of informational source that contributes to the development of living organisms (Griffiths and Gray 2001, Sterelny and Griffiths 1999).'

The problem of what carries biological information, *e.g.* of development, is shown most succinctly by citations from (Godfrey-Smith and K. Sterelny 2007): 'For many biologists, the causal role of genes should be understood in terms of their *carrying information* about their various products. That information might require the cooperation of various environmental factors before it can be 'expressed'... there is a 'parity' between the roles of environmental and genetic causes (Griffiths and Gray 1994). ... Eva Jablonka (2002) **...** treat environmental signals as having semantic information, along with genes, if they are used by the organism in an appropriate way.'

This wide problem of the carrier of biological or hereditary information acquires a new theoretical base in the discussion on the symmetry of information and code in the theory developed in this paper. Assumptions are placed outside of the area of life-referenced terms and ideas. This is a reason for using the somewhat strange term 'living object'. On this deductive way we meet a well defined process which is very similar to the natural life process. Probably this is a definition of life, but the 'river paradox' problem should be better resolved to reach such a conclusion.

McKay (1991, 2004) states: 'A simpler definition (of life) is that life is a material system that undergoes reproduction, mutation, and natural selection'. Maynard Smith and Szathmáry (1999) (who were particularly appreciative of information) also indicate reproduction and mutation but they use heredity instead of selection and add metabolism. I claim that reproduction by definition encompasses heredity which is a parameter of changeability (mutation) speed only. Selection can be derived from the limited resources of a physical environment (as Darwin suggests), but for this and metabolism or a 'material system' a physical environment must be assumed which limits theoretical attempts and artificial life especially. Up to now, we have not a good, commonly accepted definition of life (Rizzotti 1996, Taylor 1999, Lahav 1999). I will not recollect it here, and (Lahav 1999) should be sufficient.

Reproduction, Darwinian mechanism and the natural identity criterion all result from the **spontaneity** of purposeful information. While investigating artificial life, we can learn many about natural life, but we keep in mind the questions' limitation resulting from the differences between the artificial and natural origin of an investigated process. In artificial life, the basic properties are usually (but not always) assumed and we cannot ask of it in such an experiment.

I understand natural life as spontaneous, long and as soon as possibly effective process of collection of *purposeful information*. If it is life, then it should be explanatory of what is observed here and now. Such a potential definition is similar in meaning to McKay's definition, but different is 'bootstrap'. In the first step, therefore, I define the term: '*purposeful information*' using the somewhat generalized terms *code* and *information*, mainly to widen the range applicability, but still exact. These small changes of meaning of the very base notion, needed to fit new tasks in this complex jigsaw, may be the great problem for readers familiar with the commonly taken meaning. Pleas forgive me for this.

Mathematical notation is not needed to construct an exact definition, but it is useful. If you see a mathematical expression you typically expect that it is for calculating something useful. In this paper expressions are not meant to calculate, but to clarify descriptions and understanding.

## 2 Defining Purposeful Information and Its Environment

The way to define *purposeful information* (*e.g. biological information*) leads through several simpler notions which must be defined ahead of time. These are *information* and *purpose*. For *purpose,* the notions *cause* and *effect* are needed, and for these *natural code* which describes causal conversion of physical information form.

### 2.1 Generalization of Information and Coding

The definition of *purposeful information* aims to describe real physical objects which are developed by biological evolution. Information, as introduced by Shannon, is defined in the range of human technical applications. To be suitable for the description of physical objects it needs to be generalized. The dynamic aspect of information flow and conversion of its form describe the notion of *coding* which also should be expanded to include cases connected with spontaneous, causal physical transformation without human involvement.

***Information*** is a choice made from a set of available possibilities. We had been accustomed to meet information in indirect form. We usually obtain symbols, which only represent an actual choice in some set other than a set of symbols. The chain of representations ends on an object which is the semantic meaning of the considered information. *E.g.* for Maynard Smith and Szathmáry (1999) and Jablonka (2002) information is not recognisable if it is not encoded. Here, we are interested in the choice just direct in this last (semantic) set. Representation will be considered later using code.

Any ***object contains information in its structure*** how to react to external conditions – *the environment*. Example of the lack of some information: a state of non-stable equilibrium of a ball on top of another ball. Together they form a single object. The upper ball will fall if it will 'know' the direction, *i.e.* if the choice of direction would really exist.

***Code*** is any transformation in a pattern of any relation or conditional probability, which makes it possible to connect the two sets, in that the choice in the first set causes a choice in the second one. The code allows the symbolic representation (indirect choice), *i.e.* to record (encode) information in a set other than the 'last' (semantic) one. In general, we can write: $y:=f(x)$ where $x$ is a choice from $X$ which is encoded by code $f$ into its symbol $y$ from set $Y$. We use the sign ':=' like in a programming language, not '=' like in mathematics because this transformation is an action with a direction, not a static identity or equality.

### 2.2 Isolated Physical System and the Natural Code

**The *Natural code*** is a transformation by physical law in the direction of time but it must be defined more exactly: Let $u \in U$ be a state called ***situation*** of some ***isolated physical system U***. Physical law $F$ transforms $u_1 \in U_1$ (the ***cause***) to state $u_2 \in U_2$ (the ***effect***) at a next discrete time point. $u_{t+1} := F(u_t)$

Using some ***criterion*** we can find in this *situation u* an ***object o*** and the rest of this situation we call an ***environment e*** of *object o*, so $u=o+e$ . Now we can write: $u_{t+1} := (F+e_t)(o_t)$ (The sign '+' is used here in its intuitive meaning, it may be formalized.) We can forget to write $F$, because there is only one $F$ and thus it is not a choice and not a piece of information.

*E.g.*: A known chemical reaction: $Na_2O + H_2O = 2NaOH$ can be denoted in our notation as $o_1=$'$Na_2O$' ; $e_1=$ '$H_2O$' ; $u_1=o_1+e_1$ ; $u_2 := F(u_1)$ ; $u_2=$ '$2NaOH$' . In chemical notation nobody uses $F$, we also forget it and we write: $u_2 := e_1(o_1)$ or $u_2 := o_1(e_1)$ . If there is a large quantity of one substrate, (*e.g.* of water) and very few molecules of second one, then we can neglect change of first one. Let 'particles with Na' is the criterion of object: $o_1=$ '$Na_2O$' ; $e_1=$ 'large quantity of water $H_2O$' ; $u_1=o_1+e_1$ ; $u_2 := F(u_1)$ ; $u_2=o_2+e_2$ ; $o_2=$'$2NaOH$' ; $e_2=$ 'large quantity of water $H_2O$'; then we have an environment that is so similar that $e_1=e_2=e$. If we have the same $e=e_t=e_{t+1}$ then we can denote this as: $o_{t+1} := e(o_t)$ . In this case we like to call $e$ a ***natural code***.

If $o_1$ is a frog egg, $e$ is a puddle in the Spring then $o_2$ is a tadpole and $o_3$ is a small frog. If $o_1$ is a sweet juice and $e$ is a yeast, then $o_2$ is a young wine. Both parts ($o$ and $e$) of situation $u$ are still symmetrical and each of them may be a code or a code argument (***cause***) and code value (***effect***) but, for the purposes of this paper, we assume, as soon as it is possible, a stable environment $e$ as a code. Changes of object $o$ should not have any significant (observable) influence on this code. This is *an **assumption*** *of 'pass over'*. We see an object as much smaller than its environment, and the description of the environment is simplified and statistically averaged, containing in one its state many similar particular, real states. It means that states of environment are differentiate by few parameters but rest of parameters are statistically averaged. This disproportion and not precise, statistical character of environment will be later needed to hide in the environment a spare copy of object. Such environment should not be visibly changed. This copy is necessary for process continuation when first copy of object was eliminated after miss change.

### 2.3 Reverse Natural Code and Purposeful Information

***Purpose*** is an assumed *effect* for which a *cause* is sought. This search is the transformation of a known *purpose* to a sought *cause*. It is a reversed code in relation to the natural code, but it should work chronologically and be determined at least for this one, particular argument. Typically, such a reversed code does not exist - it is not a natural code. We allow for code to be multiple-valued which leads to information-losing. Most physical transformations proceed in one direction, that of increasing entropy, *i.e.* they tend to lose information during encoding. It means that particular *purpose* may be obtained in a few different ways, *i.e.* it may have few different *causes*.

***Purposeful information*** is a record of a founded *cause* coding an assumed *effect* – *purpose* (goal). It is a choice (in set of any causes of any corresponding effects) adequate to obtain a purpose. This adequate choice becomes *purposeful information* if it is checked that it leads to the *purpose*. *Purposeful information* is defined for particular, constant code, however, small change of code may lead to small loss of *purposeful information*. 'Biological information' is an example of purposeful information and the quantity of purposeful information corresponds to Eigen's 'information value'.

***Construction*** is a general algorithm (its real physical mechanism, form of its existence) which uses the *natural code* to perform its reversed code which works in a time direction.

Because, in general, the reversed code working in a time direction is not available (it is given in an implicit form) the *construction* must be constructed using the explicitly given *natural code* to *test* many ***hypotheses***. The general construction must set down these *hypotheses i.e.* to consecutively submit many of different *objects* to transformation through the *natural code*, out of which one may turn out to be the *cause* of an assumed *effect* (*purpose*). The construction must allow the *natural code* to transform each of them and, after the transformation, to compare each effect to the assumed purpose. Transformation and comparison together constitute a ***test***. To create *purposeful information*, the *construction* must also record its effect of searching, *i.e.* to record the found cause of the purpose. As can be seen, this is the base pattern of the Darwinian mechanism, but still lacks the necessity of reproduction and particular purpose: 'continuation of existence' with an object identity criterion appropriate to it .

We often use such construction when we need to find the correct *x* to have *e.g.* $8=f(x)$, but *f* is complex. It is easier to calculate *f* for few values of *x* and find one such as $8.1=f(3)$ which can be enough than convert function to reverse form.

## 3  Variants of Purposeful Information Quantity

Later, we will need the amount of the purposeful information to detect its growth. It is based on the probability of selecting the proper cause (or set of causes) for a given effect from a particular set of available choices.

### 3.1  Single-valued Code

Let us consider the single-valued (each argument is mapped to at most one value, it may be many-to-one or one-to-one transformation) code  $k \subset X \times Y$  *i.e.*:  $y := k(x)$  where $x \in X$; $y \in Y$; *x* represents a given cause, and *y* represents the effect of this cause. Lets use $g \in Y$ to indicate a particular purpose. The construction can be expressed now as "quasi-inverse" of *k*, (that may be a multiple-valued code) $x_g := k^{-1}(g)$. The selection of a cause $x_g$ gives us all the necessary information to obtain the purpose. If *k* is injection (one-to-one code), then the cause $x_g$ is single and unique. Using a typical Shannon equation we have:  $I(x_g) = -\log p(x_g)$, where $p(x)$ is the probability distribution of drawing any cause *x* .

## 3.2 Global or Local and Objective or Subjective References of p(x)

In the *situation* - a state of our system *U*, we like to see many different objects *x* simultaneously, with a certain probability distribution *p(x)*. Using a particular *criterion* we select one of them as our object and the remainder makes up the environment *k* of our object.

The ***objective*** determination of the *p(x)* distribution can be based on the specific state of the system with maximum entropy and minimum information. We use it as a reference to define the probability of the spontaneous appearance of a given object in a system. I call this case a ***global** probability distribution*. We can interpret it as a system in an *abiotic equilibrium* (Gecow 2009b).

The distribution *p(x)* derives from the question of the origin of object *x*, how difficult is it 'to make it'? We then take the system without object *x* and we ask what is the probability of it appearing in the next moment. However, depending on our interest, we can remove all living objects (this is *global* case), only particular species or objects with a particular trait.

The real mechanisms inside a given system generate the next hypothesis from the particular current state. It is the basis of a second approach to *objectively* determine *p(x)*, called **local**. The sequence $\{x_i\}: x_{i+1} := k(x_i)$ is interpreted as a process, in which the consecutive effect is treated as a hypothetical cause of a sought goal. In such a case with a single-valued code, the *objective local p(x)* is unclear:
(1) For each element there is only one following element therefore it should be $p_i(x_{i+1})=1$. (2) The given sequence usually uses only a small part of the available $x \in X$. (3) In this sense, the local *p(x)* depends on the initial starting point $x_0$. Also, the answer to the question 'how many time it must be drawn to obtain this goal' depends on $x_0$. If we ask about the *objective* probability of finding the sought goal *p(g)* or its cause in a given sequence of single-valued code, we can obtain only 1 or 0 (see 'river paradox' below, ch.4.4). We cannot base the amount of *purposeful information* on such a distribution. However, from the ***subjective*** for object $x_i$ point of view before encoding (*i.e.* processing $k(x_i)$), any *x* (*i.e.* $x_{i+1}$, as the sought cause of goal *g*) has an equal probability to be an effect of $x_i$. The need to check if a given cause really leads to goal, has some elements of such subjectivity. *Objective, local* definition of *p(x)* get a sense when we use a conditional probability $W(x_{i+1}/x_i)$ as a code. An objective existence of many possible effects (values) of such transformations may be the effect of averaging some parameters. In such a case of multiple-valued code *W* the local $p(x) = p_i(x_{i+1}) = W(x_{i+1}/x_i)$.

## 3.3 Selection of an Element or Subset

Any process of constraining the choices from *X* to $A \subset X$ is an information useful for attaining the goal, not only the selection of particular individual causes *x*. Thus, any process of elimination of any causes which do not lead to the assumed effect or lead to the goal with insufficient probability, should be treated as information.

The information quantity in such a case is equal to: $I_X(g;A) = -\log p_X(A) = -\log \Sigma_A p_X(x)$. This information I call the ***set-type information*** in contrast to the ***element-type information*** which is based on selecting a particular cause *x*. If set *A* is connected to some property of the object, then the detection of this property in a hypothetical cause gives $I_X(g;A)$ bits of the purposeful information (if we assume a logarithm with base 2).

## 3.4 Aptness in the Multiple-valued Code, Process of Improvement

A typical biological environment (such as *natural code*) of a particular population (and of each object *x* from this population) is much better described by conditional probability than single-valued dependency. In the case of multiple-valued code *W* any cause *x* encodes the goal *g* to some degree that is termed ***aptness** of x* (similar to fitness) and measured by the conditional probability *W(g/x)*. During the test, the object *x* can get an information that its *aptness* reaches a particular level (*e.g.* object *x* can survive).

The amount of purposeful information and the aptness increase together (see Gecow 2009b). The process $\{x_i\} : W(x_{i+1}/x_i)$ is called a ***process of improvement*** (or ***adaptive process***) if the aptness is not decreasing $W(g/x_i) \leq W(g/x_{i+1})$ (***adaptive*** or ***improvement condition***). Such a process is *objectively* defined only for the multiple-valued codes.

# 4 Main Properties of Spontaneous Process of The Purposeful Information Collecting

## 4.1 Decoding Construction Inside the System

Typically we encounter *constructions* which are intentionally created outside the system, inside which the *natural code* operates, and by an entity from outside of the system. Scientists design experiments in this way (including artificial life). Particularly, *purposeful information* is memorized outside the experimental system. In controlled experiments starting conditions are also defined from the outside.

Is it possible that a *construction* collecting *purposeful information* appears spontaneously?

***In the case of spontaneous constructions, everything must happen inside the system.*** This is the main assumption and we will investigate further what are the implications of this assumption.

## 4.2 'Continuation of existence' is the Only Spontaneous Purpose, Single-valued Code

Let us consider the simplest case of single-valued code. Here, the whole cause is converted into an effect. After this transformation by the natural code, only effect exists. To create purposeful information, the *construction* must record the correct cause of a purpose (in a form useful to achieving this goal in the future). Thus, the cause must be recorded in the effect. It seems that this can occur only if the cause and effect are identical. This condition (as an object property) may be taken as a purpose.

In the 'river paradox' discussed later, we find wider possibility than identity of cause and goal. More exactly, the cause must be contained in the purpose. Only then (for identity and wider case) can the purpose represent the cause in itself, when only the effect equal purpose remains. In such a case, the marking of the correct cause (hypothesis) is inherent in the process - only correct hypotheses exist, and the wrong ones simply cease to exist. This differentiates marked and unmarked objects, and leads to another set of correlated properties later used to distinguish between objects that are alive and others.

The encoding process contains all the 'duties' of the construction: it puts forward hypotheses, compares the effects with the purpose and marks the correct trials. It does it all for the invariant point of the code, which allows the construction to be embedded within the system.

Because in definition of *purposeful information* the code is given and constant, then the only carrier of *purposeful information* in the *construction* is the argument of the code - the tested object (see ch.5.3).

In conclusion, we ascertain that the only purpose of the object is to be the same before and after transformation of the natural code, *i.e. continuation of existence*. This requires the *identity criterion*. Note however, when using single-valued identity, which is the case with single-valued codes, the process will stop after the first correctly found cause. Such a sequence of negative answers of the test, ended on one positive answer, after which only duration remains, let us call ***negative process of collecting purposeful information***. It is not the case we are looking for.

## 4.3 The Process Should be Long and Uniform like Life, Multiple-valued Code

We are going to describe natural life and its evolution which leads to an increase in complexity and biological information. Based on the observed properties of natural life, **we require our process of collecting purposeful information to be long and uniform**. This is in contrast to the previously discussed case with the single-valued code, where there was only a single act of finding the correct encoding after which the process becomes stationary. The new requirement means that the process of collecting information will be a sequence of positive answers of the test, therefore let us call it ***positive process of collecting purposeful information***.

These assumptions suggest that the *identity criterion* cannot be exact (single-valued), it musts allow changes necessary to the collection of information. The increase of purposeful information can be realized on two ways:

**(1) The goal is achieved fully** and stays at this maximal level, but probability $p(x_i)$ decreases, *i.e. element-type, global purposeful information* grows. This is a case of ***horizontal diffusion***.

Later (ch.5.1) we will consider the *negative process of collecting purposeful information* which finds reproduction at a sufficient level. From the time of this finding, *horizontal diffusion* begins. But also a *positive process* of *collecting purposeful information* begins during the so-called quantity explosion (ch.5.2).

**(2) The goal is achieved not fully** and can be achieved better. Now, *set-type purposeful information* should grow. How is this possible?

Information about the sequence of verified elements and the current point in the process must be contained in the object. One parameter is enough, *e.g. aptness,* which should not decrease (*adaptive condition*).

(2.1) **If** such a parameter is used in the code mechanism to calculate the next element of a process (*i.e.* the next hypothesis), in the way: objects with an aptness below the current level do not belong to set of possible hypothesis, **then** *construction* is *objectively* not needed – each hypothesis is accepted ('*directed process*' case). But *subjectively,* the object does not know what will happen and to create *purposeful information,* the object checks it using the *construction*. Probably this dual interpretation creates a 'river paradox'.

(2.2) **Else** aptness must be always (in *objective* and *subjective* cases) controlled by construction's elements: test and elimination (together - selection), which can break and end the process of collecting information. It is obvious that to lengthen such a process reproduction is needed, but we will discuss this later. In such a case, the aptness parameter for the goal '*continuation of existence*' is called 'fitness' - a measure of how accurately the purpose of '*continuation of existence*' is encoded. This leads to the multiple-valued, conditional probability-based code *W*.

### 4.4 Directed Processes, 'River Paradox'

The requirement for code and objects to have special properties which prevent to put worse hypotheses is very strong, but we can show such physical processes. For example, any predictable physical processes leading to some state, such as coming to thermodynamic equilibrium. At each time step, an object is nearer equilibrium, which can be treated as its goal, and therefore has more purposeful information about this particular goal. These processes (I call them ***directed processes***) look like single-valued sequences where the limit is defined at the beginning: a choice in the set of available possibilities was *objectively* made. As was mentioned above, the necessity of *construction* depends on the type of purposeful information in this case. For *objective purposeful information* it is not necessary, but for the *subjective* one it is used.

Much harder to understand is the next example: the '***river paradox***'. On a new big, dry continent (it may be a labyrinth) in one place there is a source of water. Water seeks the way to a sink (sea or exit from the labyrinth). Water always will find a way to the sink and a base current will flow along the shortest possible path.

It makes many 'false' hypotheses before it finds this way, but they are not eliminated, only memorized in later objects to not be checked a second time (see ch.4.2: the cause must be contained in the purpose). The found path is recorded and this record is later used. This creates canyons. Here, it seems that it is 'collecting of purposeful information', however this is also an example of '*directed processes*' with the effect being defined at the beginning. There is no choice to make.

The pseudo-randomness of the process creates serious problems of interpretation, together with a base question of type of determinism in nature and a connected type of randomness in natural life. A broad and deep discussion on such problems and a better understanding of the place of *subjectivity* (in the above defined sense) is needed, but it will only be possible after recognizing their environment, which is the aim of this paper. Because the breadth of the investigated questions together creating an attractive path, this paper can be only a simple draft of this path. Probably, this is the only problem which must be solved before we can agree that a 'spontaneous, long, uniform (and effective) process of collecting purposeful information' is the definition of natural life which we see around.

### 4.5 Small Change Tendency, Natural Identity Criterion and Damage Avalanche in Chaotic Systems

It is an obvious simplification that a particular ideal object *g* exists as a goal for living objects but in the first approximation for reconnaissance it is useful. The aptness $W(g/x)$ of the object in the internal construction of the system can be treated as its probability to exist further. The aptness *W* is an effect of object building. We may expect that smaller changes of building typically lead to smaller change of aptness and purpose information quantity. In biology, the term fitness is used for such aptness. Based on this, we can define a distance $D(x_1,x_2) = |W(g/x_1)-W(g/x_2)|$. (For $W(g/g)$ we take *max* $W(g/x)$.) Note, that changes (*i.e.* $x_i$ into $x_{i+1}$) greater then the distance to the purpose $g$ (*i.e.* $D(x_{i+1},x_t)>D(g,x_t)$) must be eliminated by the adaptive condition $W(g/x_{i+1}) \geq W(g/x_i)$. In addition, in the range of this distance, greater changes fulfil the adaptive condition less often (Gecow 2009b). This important **small change tendency** inherently creates a **natural identity criterion** which requires the absence of large change in the sequence of the evolution of 'the same object'. But small changes are allowed, therefore it is multiple-valued criterion of identity. For collecting purposeful information, it is crucial to lose less of it than to acquire new when much is already collected. If more than one bit is collected, then the probability of loss must be greater than finding new information and a larger change must be statistically more dangerous (in that it decreases aptness).

What does 'large' or 'small' change mean in *small change tendency* or *natural identity criterion*? The threshold can be objective. It is derived from 'damage size distribution' in complex systems (Gecow 2010a,b,2011,2008a). This is not right place to describe the differences between ordered and chaotic complex networks, but a few simple words should be written. If we have two identical functioning Kauffman networks and in one of them we introduce a small disturbance, then the subsequent processes will differ. This difference is termed 'damage' and can be measured. The distribution of damage size depends on network parameters. The network is chaotic when most small disturbances lead to large damage avalanches, after which the network functions in an absolutely different way. For chaotic networks, there are two peaks in damage size. For very small damage, which fades out after few steps, and for very large damage, but intermediate damage is lacking. Biological evolution, as our collecting of purposeful information, takes place only in the area of very small changes (damage which fades out). A large damage avalanche means death, *i.e.* elimination by adaptive test, because the natural identity criterion is not fulfilled. This view is so different from Kauffman's 'live on the edge of chaos' hypothesis, that it practically is contradictory to it. However, I also place live near edge of chaos, but in a much more chaotic area, on the chaotic shore of Kauffman's 'liquid area'.

The measure of achievement of the purpose *continuation of existence* is equal to the probability of keeping one's identity intact. The *natural identity criterion* must allow some changes to hypotheses can be set. Because $y=x$ in natural criterion exactness, it is sufficient in most cases to write $W(y)$ or $W(x)$ instead $W(y/x)$ for aptness.

The test based on the *natural identity criterion* applies to a particular event, which is later averaged to calculate aptness $W(y)$ to compare it to the presently required level. A mechanism of this comparison and level growth will be discussed later. Changes in the range of the *natural identity criterion* may improve or worsen the aptness of an object. Note that the natural identity criterion describe inheritance so important in the definition of life proposed by Maynard Smith and Szathmáry (1999), yet we still do not have reproduction. We still do not have death of old age, because we do not consider sexual crossing (see ch.5.8). One object evolves. Later one object will comprise a chain of all the ancestors of a given object seen in particular time point. Reproduction by definition means making more identical objects. In our interpretation (*natural identity criterion*) they are the same objects as their ancestor. Reproduction needs no additional assumption of inheritance to describe possibilities and their limitation to change. These possibilities and limitations lie in the natural identity criterion derived from earlier assumptions of information collecting.

### 4.6 Controlled by the Purposeful Information and Random Parts of Changeability

According to the purposeful information collected in an object, the transformations by code *W* should maintain an object's identity, which happens successfully with probability $W(y)$. It is a source of stability. A sequence of such changes (transformations) controlled by collected purposeful information can be interpreted as, for example, life cycle (see ch.5.2), but not mutations (new hypotheses).

Transformations by code *W* should also put forth random hypotheses *i.e.* random changes of an object. 'Random' means that the effect (direction aspect) is not controlled by the purposeful information. Such purposeful control may concern only parameters, *e.g.* like size, time and place of changes or frequency distributions. This mixed type of random changeability with a factor of purposefulness (*i.e.*

connected to needs) creates a 'Lamarckian dimension of evolution' (Jablonka & Lamb 1989,1995, 2005), see ch. 5.1.

Random changes can, in the next step of time, increase or decrease the aptness of object, *i.e.* add new or lose purposeful information. This part of the transformations corresponds with the mutations. Many of them will not fit the *natural identity criterion* and terminate the process. Elimination by *adaptive condition* (which mechanism is not yet discussed) should additionally eliminate changes that decrease aptness but are small.

Both components of changeability are different. When we are interested in one of them, we can neglect the second one as was done in the model described in (Gecow 2009b) which studies random changeability. An investigation of this random changeability will be easier if we clearly detach it from the controlled portion. We are interested only in the effect of change – if it does not decrease aptness $W(y)$. The cause of this change is not more interesting. In this case, we lose one of the main aspects of purposeful information that indicates a cause of the assumed effect. However, it was already used to hold the *construction* in the system and to indicate the goal: '*continuation of existence*'.

The next simplification is treat aptness only as the distance to the goal (not a probability of existence). It frees a change size distribution from the range allowed by the *natural identity criterion*. This simplification makes it possible to see the basic cause of a *small change tendency*, which is the base of the *natural identity criterion* and the base of a few other important tendencies of changeability termed structural tendencies.

These two simplifications are a base for the model described in (Gecow 2009b). There, object *y* is compared to its ideal *y\** (goal). The code *e* ($y:=e(x)$, *e* is an environment of objects *x* and *y*) is constant by assumption. Because of the *natural identity criterion*, practically *x=y*, and thus *x* and code *e* are ignored (not interesting), and only random changes of *y* are investigated in the model. Using this extremely simple model we can get accustomed to properties of aptness and distribution of fulfilling an *adaptive condition* for different parameters.

After the displacement of the object and its environment in such a way that an active code becomes an object, there appears the possibility of investigating the changeability of a functioning structure of object. A directed network (such as a Kauffman network) called RSN (Gecow 2010a, 2011) is a model of the object. The changes of structure are exposed on the *adaptive condition* by result *y* of structure function (in the way set out in (Gecow 2009b)). In this model, structural tendencies of changeability are investigated (Gecow et al. 2005, Gecow 2008a, 2009a).

### 4.7 Three Causes of Reproduction

The main duties of a *construction* are: to compare an object and the purpose, and to record an effect of this comparison. In the *improvement process* it is to fulfil adaptive condition $W(y) \geq W(x)$.

(1) Just a measurement of statistical aptness $W(y)$ – the probability of survival, requires the *reproduction* of an object and time for observation. This measurement concerns a population of identical objects, not only a single object – this shows what we investigate. (This is not a population based on sexual crossing as mentioned in ch.5.8.)

The demand for the process of collecting purposeful information is *long* practically requires *reproduction* due to the need to compensate for the loss effected by two independent causes (connected with the two above defined portions of changeability);

(2) The probability of 'death' $1-W(y)>0$ because the object is not ideal;

Let us introduce $V(y)=1/W(y)$ - as the *reproductive rate* required for this compensation.

(3) The elimination of erroneous hypotheses. Objects changed so as to find more purposeful information.

Let *N* be the number of objects in a population and $B=N_2-N_1$ be *balance*. For a large *N*, one can define a *sufficient reproductive rate* - *v* for balance $B=0$ over a longer time, which compensate both causes of loss: 'death' because the object is not ideal and the elimination of erroneous hypotheses.

Reproduction only multiply the number of identical objects (without changes). Change, interpreted as mutation which is the source of new hypotheses, is a separate action. In this paper (excluding ch.5.8), we do not consider information interchange, *i.e.* crossing, and a parent is single. After reproduction, only objects of the next generation exist. Heredity, as understood in (Maynard Smith and Szathmáry 1999, Taylor 1999), is exactly included in the general term 'reproduction' which is based on the natural identity criterion.

Indicated above three causes of reproduction concern the object which collects purposeful information, *i.e.* the object which is tested and may be eliminated. Existing for these reasons reproduction but absorbed by object *e.g.* for soma construction (ch.5.5) has radically other condition which should be taken under consideration when heredity is analysed.

## 5  Hereditary Information

Reproduction, especially in the context of biological information, makes think of heredity and hereditary information. However, hereditary information is not a well defined term and it is usually understood in extreme ways, such as genetic information. The way shown here is developed as deductive draft. It defines its own terms to clearly separate their effects which can lead to meanings different than traditional ones. As mentioned above, reproduction does not require any assumptions about heredity. This reproduction, however, is simple. One parent transforms itself into several identical and 'the same' descendants. It can be found in the living world among procaryota, but even there horizontal information exchange also occurs. Eucaryota usually have much more complicated development with a robust interchange of information in the crossing-over process. It will be discussed in the next papers of this draft, but the conclusion of hereditary information without crossing-over is too simple for application and some corrections are already needed and added in ch.5.8.

### 5.1  Object Transformation Cycle, Hereditary Information as Purposeful Information

Now, let us assume that there is no random component of changeability. In effect, we will consider only that portion of changeability controlled by purposeful information. Let us neglect the fact that mutations and evolution are necessary for long-term existence. We can ask: What changeability remains, if purposeful information should lead to keeping the same object in the next time step? It means that there should be no changes.

Really, the lack of changes is one possibility. The second possibility is a *cycle* of transformations which is generally met in practice. In this case, the time step which we have used is the period of the *cycle*. After the period, the object (current form of the cycle) is the same. But the cycle consists of a few smaller time steps (transformations), and objects (their arguments and results). Each of the objects may be different but it must contain all *purposeful information* which keeps the *cycle* stable and indicates all transformations and objects of the *cycle*. Such a description exhibits aspects of 'the developmental system (DS) view of Gottlieb' (Jablonka 2007).

The *cycle* can be treated as a finite chain of objects, like above. We can take a whole cycle (or the state of a cycle in a particular phase) as an object in the old meaning, where the object does not change its identity and form. Natural identity criterion concern this view. Now however, we are going to describe such cycle in more details. Taking that one object exists during whole cycle and later (even with mutations and reproduction) in next cycles is the third possibility. In such a case it changes *form* but identity remains. This view is the easiest to use. This meaning is used in term: 'object evolution'.

Another *object form* is another *form* of *purposeful information*, but semantic aspect of this information does not change. Meaning is still the same - the whole particular sequence of *object forms* constituting *cycle*, as well as the object in third interpretation. I call this constant *purposeful information* in the *cycle*, independent of any current *form* - ***cycle information***.

For definition of *cycle information* we do not use reproduction. Cycle of transformation is possible but not necessary. It can be a constant sequence which can be split into formal cycles. But in physical conditions the reproduction (which necessity was shown in previous chapter) needs collection of matter for growth after reproduction. This needs a cycle with different forms of object. In such cycle a special time point exists - this is a first time point when there are more than one object. In practice it may be a large period up to stop of interaction. The transfer of *cycle information* through this special point/period to the 'new' (*i.e.* descendant) object is called **inheritance** and transferred *cycle information* is called **hereditary information**. If all objects after reproduction are so similar that is no way to indicate 'old one', then we can treat all of them as descendant.

For inheritance problem a mother object (before reproduction) and descendant objects are different (not the same) objects but for evolution investigation in long term they can be treated as the same object especially, for a'posteriori view.

In current stage of model development there is no base for consideration and gauge of various alternative properties of object - object is tested as a whole. For observer which can compare independent objects a *cycle information* and *hereditary information* can curry more information than the *purposeful information* needs (see ch.5.8).

Note that above view does not openly assume a single-valued code, which is intuitively the simplest case. Using the more adequate code *W*, this view becomes much more complicated but more adequate. Described reality is much more complex. In one life cycle can be a few, even different type, reproduction events. Other problem appears when environmentally induced regulative object reaction is longer than cycle. Inducting change of environment can be treated as long fluctuation, however, in the range of stability of statistical environment. In such a case the *hereditary information* curries specyfic state of this regulation mechanism, correspondent object forms in consecutive cycles are different. It can incorrectly suggest that purposeful-, cycle- and hereditary information is changed in comparison with typical state and evolutionary change occurs. This problem, however, is the same like for different *object forms* in *cycle*, but now the cycle closes when environmental fluctuation disappears. Jablonka & Lamb (2005, 'Self-sustaining loops') and Jablonka (2007, considerning notions 'plasticity' and 'canalisation') consider such long purposeful reactions in context of heredity and evolution. A change of *purposeful information* should be taken as evolutionary change. It is connected to 'units of evolutionary change' sought by Jablonka (2007). By definition, it can appear only in effect of random change testing. This randomness describes a choice not supported by already collected *purposeful information*. Any purposeful modification (using collected *purposeful information*) of possibility set or probability distribution of such choice creates 'Lamarckian dimmension of evolution' (Jablonka & Lamb 2005). Local approach to purposeful information (ch.3), especially using code *W*, gives teoretical support for deeper understanding of this problem.

## 5.2    Code Cycle, Environmental Changeability over Time

Up till now, we have assumed a stable environment, *i.e.* constant code. It was necessary to obtain the same effect if the same cause was used. In the same way as for the object above, this assumption was maintained for the whole time step, *i.e.* for the period of the object's cycle. This now means that each *object form* in the cycle has the same code as in the same phase of the previous cycle exactly one period ago, and the code can also be a cycle. Such a code cycle may be the source of the object cycle, but not *vice versa*. This is form of the *assumption of pass over*. In this way, we make second step to introduce changeability of the environment in a temporal dimension. First step has been a fluctuation.

If synchronization is broken, then although seemingly nothing has changed in the object or in the code, but all purposeful information usually disappears. We did not assume similarity between neighbouring codes. It cannot happen in single-valued code, but we know that this is only an approximation. More adequate code *W* leads to too complex view which, in order to be partially understood, needs such an approximation.

## 5.3    Environment Changeability in Space, Code Indication

Up till now, only a single, given code *e* - environment has existed in one time step. An object (or object form) has no choice (for code indication *e.g.* by code transformation) because of the *assumption of pass over*. We know, however, that in the reality which we are going to describe there are simultaneously exist many different sub-environment and using only one is a major simplification. Similarly, we can describe spatial heterogeneity of one environment. It means, that we need to split our full environment (which remain in situation *u* after removal of the object *o*) into lot of sub-environments. Object is in one of such sub-environment which play role of code like previously: $o_2:=e(o_1)$ . Let's now use for full environment letter ***E*** as a set of sub-environments *e*. Object makes a

choice - it indicates its sub-environment $e_0$ in the $E$ but it does not transform $E$ nor any $e$ therefore the *assumption of pass over* is fulfilled.

Let's consider an example: Let flat space of $E$ is a lattice *e.g.* 10*10 of equally probable $e_{ij}$. Then choice of particular $e_0$ by object gives $I=log_2 (100)$ bits of information. This choice is a property of object. Let object with equal probability can migrate to 8 neighbouring $e$ and $e_0$, but it survives only in $e_0$. Its purpose information limits its movement to 9/100 of possibilities and is $I=log_2 (100/9)$ bits but should limit it to 1/100. Reproduction compensates incorrect moves. In this view the limitation of $E$ to 100 elements plays crucial role, however, in description of real 'separate system' $U$ and particular object $o$ such limitation may have no objective bases. In more 'local' description for particular object with its choice of $e_0$ only 9 elements have not neglectible probability and boundary of $E$ is not important. In such a case object has no purposeful information, but if object changes its properties and this way probability to stay in $e_0$ increases, then purposeful information will appear. Maximum of this information is $I=log_2 9$ when only $e_0$ becomes available.

The choice (even passive move in space) on a boundary between neighbouring environments can depend on object properties and thus can carry purposeful and hereditary information. The environment $E$ and any sub-environments $e$ remain without change; the *assumption of pass over* does not concern the object move.

For example, the Atlantic Salmons return to the precise freshwater tributary in which they were born (Shearer 1992). In this case, 'the only' (see below) carriers of place indicating hereditary information are: presence of eggs in a particular place or path memory, depending on the cycle stage. Summarizing: choice of sub-environment can be a carrier of purpose- and hereditary information. More examples can be found in Jablonka & Lamb (2005) and Jablonka & Raz (2007) of behavioural heredity, where environment (more precisely: sub-environment) is subject to choice.

### 5.4 Range of Interactions, Object Influence on Environment Element - Cover

In ch.5.2, it was remarked that we did not assume similarity of neighbouring environments. However, they usually are similar in the described reality, and if they become changed, then the change may be small or large. It means that in other code the purposeful information must not be fully lost. Tools to describe and to compare environments in more detail can be based on the remark that a **particular environment consists of many factors** which actively influence an object. Similar factors can be elements of other environments. Generally, an environment can be treated as particular subset of possible active factors.

Another important phenomenon is the **spatial range of the factors influence**.

Basing on these new aspects, we can consider that an object get an influence on the presence of certain elements in the environment. It looks like a violation of the *pass over assumption*, but the element stops belonging to an environment and becomes a part of an object. It does this locally only and does not change the general parameters of a large environment. Such an influence can enlarge fitness, and thus may constitute purposeful and hereditary information as well. In this way, the environmental element is absorbed by an object. It becomes a new element of the object. I call such element a '***cover***' (Gecow et al. 2005) because this element saves the object from new, undesirable environmental factors. A structural tendency, '*covering*', was stated using a computer simulation. Structural tendencies (see ch.4.6 and Gecow 2009a) are modelled using a functioning network for object description.

In most of this article, the object is an inactive argument and the code is a function, the active part of the considered system. But this choice is arbitrary and is an effect of the lack of a better model, where both parts are similar. While *covering,* the element of code is shifted to the object. This shift changes the activity of the element, although it is not a real change, but only a simplification needed in the model. Environmental elements are simultaneously active. Kauffman networks describe them well, but also in this case there is a problem in interpretatively separating passive signals, active nodes and the developed network structure. In described nature they are similar. Process algebra, especially pi-calculus completed by module detection method (Nowostawski & Gecow 2011) is a promising tool to describe environment and object, even several objects, which actively construct their structure on one space and similar level of activity of objects and environment. .

### 5.5 Cover Levels, Hereditary Channels

Treating the cover as an active network node, we can consider a link from the cover to the object. From this link, the object gets a *covered signal* essential to the correct activity of an object. The covered signal is a cause of the need to cover when the signal disappears in object independent environment. Before covering, it was an environmental signal, thus it cannot be a carrier of *purposeful information*. Now, the **old form** of the object (I call it *core*) functions without changes because it still receives the same signal. But the *cover* is no longer an environmental element. It does not exist without the object, the test and elimination already concern it. Thus, the same signal becomes *purposeful information* despite the fact that it changes nothing. As the environment before, now **the cover seemingly does not purposefully influence structure and function of the core**. Therefore, treating a *covered signal* in such a different way, *purposeful information* which indicates the structure of *core* is carried only by the *core* (but in an old environment rebuilt by the *cover*).

The *core*, however, has been changed so much that it is necessary to keep the *cover* present. Such a change can be a node addition to *core*. Also, the *cover* is an additional node. It has to be not identical to the old environmental mechanism which has disappeared. Thus, the *cover* can maintain its existence, *i.e.* it can carry *hereditary information* about its structure and function.

If the *cover* exists in each stage of the cycle, then there are two **channels of hereditary information**: the old one (*core*), and the new one (*cover*). Further covering of an already covered object creates consecutive levels of cover, *i.e.* higher channels of heredity. If the *core* is a genetic channel, then higher channels are epigenetic. Jablonka & Lamb (2005) and Jablonka & Raz (2009) show existence and importance of epigenetic channels, but they do not use the concept of covering.

If cover does not stay during whole cycle, which can be expected in early stage of new level creation, then core must curry the purposeful information about cover. Using a part of sister objects is one of the simplest and ever ready possibility to obtain by object controlled elements of environment needed for cover building. Resignation by such sisters from unlimited own reproduction which clashes with cover role, significantly changes their status. They still can be considered as evolving objects, see ch.7.4, but such perspective becomes unnatural and not simple. This way soma and workers of social insects and even letters of plants have evolved.

### 5.6 Which Channel Carries Hereditary Information About Bird Migration

The Passeriformes have a genetically recorded migration pattern. Undertaking the journey and the choice of migration route at the time of this choice is recorded in two separated levels: genetic and phenotypic. Are any of them complete? If so, which one constitutes hereditary information?

First of all, hereditary information is unitary, and only one choice will occur, but it can be recorded in two forms simultaneously. At the stage of environmental choice, the phenotype is an active form which makes a choice and realises the journey. At this time, the genetic form is passive but destroying it results that the next iteration of the cycle does not take place. The hereditary information should, by definition, contain all that is necessary to realise the next cycles. It is not necessarily possible to indicate a carrier of a shown property in a dynamically changed object. It is, rather, possible that whole object carries such information (see ch.5.8).

It has been possible to teach a migration route to a flock of Anseriformes such as geese. Here, the main part of this hereditary information is carried by the behavioural channel placed in the tradition of a flock. Jablonka & Lamb (2005) and Jablonka & Raz (2009) prove the existence and importance of such epigenetic channels. However, hereditary information carried by this channel is also incomplete. It needs the full object and probably a flock to reach the same stage of the cycle a subsequent time.

It is usually taken that the genetic form of hereditary information fully indicates the phenotypic form. But it is not sufficient - abandoned in an environment it will perish. To constitute information about migration which directly makes a choice, it must be decoded to the phenotypic form. The decoding is performed by all remaining part of the object including mother, father and even the nest. (The environment independence of the object is the boundary.) In a majority cases, it is a cover with its mechanisms of heredity which are needed to reach the phenotypic form.

The misleading aspect of the title question, which suggests that a passive record exists and contains all needed information, results from the oblivion of the active code. Typically, peoples meet an information record and construct their intuitive projection while reading or listen. Passive information is encoded in symbols and should contain all that is needed for use by the receiver. The active code, however, which convert symbols into their semantic meaning is always present in the receiver, who himself is the converter. We also forget that information and code are fully symmetric, their passive or active form is only an arbitrary choice for simplified description. Such a habit is adequate for information between persons, but we inadequately assume that a part of hereditary information looking more passive is complete or dominant.

The need to use the whole object form as hereditary information creates a very murky problem, wherein information about a particular property is recorded. Such a place can be defined approximately, but never exactly. This problem becomes more clear when crossing-over is introduced - see ch.5.8.

### 5.7 Autonomous Model, Genome Decoding into Phenotype

Due to the effect of covering, environmental influence on the decoding process of a genome into the phenotype is small. Let us neglect it in the first approximation. We obtain a description of the decoding similar to the natural code. The object is an isolated physical system. Such system devoid of external inputs and outputs is called 'autonomous'. Its states named in ch.2 *situation $u_i$* correspond now to *object form, i.e.* i-th stage of cycle.

Kauffman (1992) considers such an autonomous system. He describes it using limited deterministic Boolean network. In the limited network there is limited, however - large, number of states of the network, therefore in limited time a state must appear second time. From this point of the time it will appear in cyclic way ad infinitum. Such cycle is called 'attractor'. Using such model Kauffman describes living cells, especially cells of soma. He shows, that for certain parameters (on the edge of chaos) of random networks a cycle length typically is not large. This view seems to be similar to our. He suggests, that we should await such small cycle in living objects and it is effect of spontaneous order appearing on the edge of chaos but I have firmly opposite opinion that I have proved in (Gecow 2011, 2010a, 2010b). First of all, living objects cannot be a random network and cannot stay 'on the edge of chaos'. They should be clearly deeper in chaos, however, still in Kauffman's 'liquid region' near edge of chaos. They are especially not random, but carefully selected by natural selection for stability, therefore their stability is not an effect of spontaneous order. Also assumption that they are autonomous is too large simplification which create ability of small length of attractor.

Let's come back to genome decoding. State of autonomous system which describes the object at a particular stage of the cycle contain the *core* which is a genome treated as passive and constant, and a *cover*, which can contain many levels. The parts of the cover elements which interact with the neglected environment correspond to the phenotype. For simplicity let us assume that each cover level exists at any cycle stage, perhaps in different form. Let us also consider a single cover level at the beginning.

We start from the zygote and stop just prior to next zygote. At each step, a new form of cover is the effect of the genome and cover form from one step before. What is the cover of the genome of a bird at the zygote stage? It is the whole zygote cell excluding only the genome. The next level of the zygotic cover is the mother. When the egg appears in the nest in one of the next cycle stages, there are several egg covers, such as the membrane and eggshell, but also the mother and father heating, as well as other protection including the nest. Most of them are built without using current genome in the egg. All these elements of the environment of an egg genome are purposeful information, and together as an entity they carry hereditary information. It is because the entity is tested, not the elements of this view. From this entity nothing can be removed without changing (typically - decreasing) the fitness - probability of cycle revolution. In other words - without changing quantity of purposeful and hereditary information. The entity can be eliminated and get one fitness as a test result.

You may have a doubt - in the first model with object independent environment also environment cannot be removed without fitness change, but only object curries purposeful information. - Yes, because only object is chosen and tested, however, in particular given environment.

This is true for simple vegetative reproduction, but above all the examples concern reproduction with a crossing-over mechanism. It has important meaning for the title question and last argument of the tested entity.

### 5.8 Crossing-Over Creates the Allele Test - the Second Test Level With Other Connections

The theme of purposeful information interchange between two objects realized by crossing-over is broad and cannot be developed within the scope of this article. However, it is important for the conclusion of the previous chapter thus should be briefly explained. Information interchange creates a new, stronger test mechanism which works simultaneously with the main test of object as entity described above, and is based on it as well. This new test concerns alternative parts of an object called alleles which can be exchanged as purposeful information carriers. Alleles are tested in this test mechanism. In effect, they change their frequency and can be eliminated. Objects are eliminated and have fitness. But population evolves, and not the object itself. The allele test can work if alleles entail alternative properties and lead to different fitness levels. In such a case, the elements of an object can be indicated that play the main role in an object's hereditary information connected with particular object properties. In contrast with the earlier conclusion, they are tested individually. They can be treated as objects with their own hereditary information which constitute the basis for Dawkins' (1976) 'selfish gene'.

The appearance of element interchange is the effect of a structural tendency of integration (Gecow et al. 2005) leading to stronger and stronger interaction of initially independent objects which evolve together over a long time. During such integration, the new common object emerges at a higher level. We consider this higher level object and its test and fitness. Initially independent objects become interchanging elements whose test and fitness are an effect of test of higher level object.

In this paper, however, we consider only simple one-level test without interchange of elements and connected to them information. Here there is no way to define and asses alternative properties of object for real process. If we take new mutation which decreases fitness, but starting balance $B$ (ch.4.7) is so high that it is still $B>0$ after change, then this change will not be eliminated, however, with information interchange it always will be removed from population. Therefore without interchange there is no 'negative properties' and if such mutated objects pass tests, then they are, nevertheless, the *purposeful information*.

Inheritance theory in that more complex two-level test should be based on reconnaissance in simpler case described here. Note, that particular mechanism of purposeful information interchange considers a particular channel of transmission of this information.

### 6 Are Obtained Properties Sufficient for the Purposeful Information Growth ?

Above, we found the main properties of a spontaneous *construction* which is held inside the system and collects *purposeful information* in a long and uniform process. First of all these are a goal – *continuation of existence*, multiple-valued natural code $W$, the *natural identity criterion*, and reproduction.

Now, we are going to check whether these properties are sufficient to enlarge the quantity of purposeful information, or in other words to create real mechanisms to enforce the process of improvement.

#### 6.1 Horizontal Diffusion

*Sufficient reproductive rate* - $v$, (which compensates both causes of loss - see ch.4.7), can be treated as the sought purpose: the first, basal and, in principle, complete, *purposeful information*. It is found through a *negative process of collecting purposeful information*, (see ch.4.2) after which there is only its maintenance. Objects which lose $v$ (*i.e.* $v$ decreases) are eliminated on $N=0$ by the basic Darwinian mechanism.

*Sufficient reproductive rate* - $v$ deprives an object of convergence to global $p(x)$ based on maximum entropy. Small changes in the range of the *natural identity criterion* allow an object $x$ to achieve points of $X$ space, where $p(x)$ is practically zero but $p(x|v)>0$ to a significant degree. This means very large *global element-type purposeful information*, which is $I(x) = -\log p(x)$. This surprising phenomenon, called *horizontal diffusion* (see ch.4.3), has an important interpretative meaning. Diffusion is random. It does not need to increase *purposeful information* in a particular step, but the probability of meeting an object with very high value of purposeful information systematically increases in a range of objects with $v$ that participate in horizontal diffusion. A higher value of purposeful information does not mean (like in the improvement process) that goal is better achieved. It is fully achieved from the outset. 'Fully' means in the first approximation only, because the assumption of *sufficient reproductive rate v* is as true as the number of elements possessed by a process (larger *N*). For more elements, however, we must wait and the rate must be somewhat higher than sufficient (*B>0*).

### 6.2 Weak Improvement During Quantity Explosion from the Effectiveness Postulate

As can be seen, *sufficient reproductive rate - v* is a fully achieved goal only in the first approximation. In the second approximation, the goal *continuation of existence* can be better achieved if *N* is larger or if *B* is larger but these are attributes of a population, not of a single object. They can be taken as parameters of aptness, but they are not the aptness which we are looking for. Let us call them *weak aptness*, and such an improvement process a *weak improvement*. Let us assume that *weak aptness* $W_0$ is related for a population to the number *N*, but for a single object ($W_1(x)$) it is related to the *individual balance = B/N* in statistical sense as probability implied from properties of object.

Any improvement process requires a *positive process of collecting purposeful information*, (see ch.4.3). We require the process to be as **effective** as is possible, such that it will lead fastest to the higher values of aptness, because we are going to compare the effects of this process to the most advanced objects now living. This condition is the basis for sorting during the *quantity explosion* which is present after the system exceeds a particular threshold of reproductive rate *v*, *i.e.* a balance of *B>0*. The sorting mechanism works through the faster growth of a population number *N* of objects of higher $W_1$ aptness. It yields a permanent increase in the purposeful information quantity. However, it may be purely due the increase in the reproductive rate *V*.

The mechanism of *weak improvements* works during the *quantity explosion*, and what is surprising, it does not use down-delimiting Darwinian elimination based on *N=0*, but up-sorting that indicates the most advanced subjects, not threatened by annihilation. However, both these type of selection are included in the notion of Darwinian selection, which is a statistical conception based on the frequency of occurrence. Selection based on *N=0* elimination, like single-valued natural code and single-valued identity criterion, is intuitively simpler, but less adequate.

The state of the *quantity explosion* is extreme and, in practice, short. It is terminated by competition as we will discuss below and then the typical form of eliminative selection based on *N=0* is reinstated. The assumption of effectiveness is for a process, competition from capacity barrier is for the environment, they make similar effects, but competition needs stronger assumptions.

### 6.3 Environment Capacity Barriers produce Competition

To stop a *quantity explosion* a limitation on *N* at the **environment capacity barrier** is required, *i.e. B=0*. When the population reaches the capacity limits of the environment, its weak aptness $W_1$ (individual balance *B/N*) suddenly decreases to zero (on average). In the first approximation, we can assume that the *environment capacity barrier* has an equal influence on all objects of population, *i.e.* $B_2(x_i)=B_1(x_i)-<B_1>$. Therefore, it practically eliminates all the slower entities in this race, those with a smaller $B_1(x_i)$ for sub-population *i* of identical objects and therefore smaller $N(x_1)$ and now $B_2(x_i)<0$. This is **competition**, it just uses *N=0* test.

### 6.4 Strong Improvement Yet Without a Known Mechanism

If random changes do not occur and the object is ideal, then $N_2=N_1V(x)$. However, for a real object $N_2=(N_1V)W$, then $B=N_1(VW-1)$. *B>0* requires *VW>1*, then *W>1/V*. We get the condition for aptness *W* defined by a particular measurable parameter. Thus, the defined aptness *W* grows for a constant *N* when *V* decreases. We actually observe such a phenomenon in biology and it fits the interpretative expectation much better than *weak aptness*. I call this kind of *improvement* **strong** and I depict it as $W_2$. In this case a reproduction together with *sufficient reproductive rate - v* is treated as the mechanism of the *construction*, not like *weak improvement*.

### 6.5 Seeming Strong Improvement Creating Complexity, Evolutionary Progress

The *strong improvement i.e.* an increase of $W_2$, appears as a result of overcoming the *environment capacity barrier*, because an object stops competing and its *N* increases without changing *V*. The next barrier decreases the quantity of real purposeful information related to strong aptness $W_2$ (which also decreases) but it is not a change of object which is the carrier of information. Nothing changed in object, therefore *seemingly* for the object its aptness does not decrease. This is a change of the environment, *i.e.* code, which we assume to be constant, and the object ceases to be reliable in it. The **length of the purposeful information record** in an object structure does not decrease, it is closely connected with object **complexity**. A discussion of such complexity needs a much more complex model, the aforementioned RSN, which is described in (Gecow 2010a,b, 2011). Also Chaitin (1979, 1987) considers connection of *length of information record* and complexity.

Strong and weak improvement, length of purposeful information record as complexity and **seeming strong improvement** - all these notions are strongly connected to the old and broad theme of evolutionary progress. 'The notion of progress is poorly defined, it can be used in a variety of different ways' wrote Taylor (1999), and the same can be derived from (Nitecki 1988). Gould, *e.g.*, does not see progress (Nitecki 1988), he awaits it in our objective, *strong improvement*.

### 6.6 The Physical Aspect of the Environment Capacity Barrier

Both the placement under control of a quantity explosion and overcoming of the environment capacity barrier require the physical, material nature of environments and objects.
Such limitations are not observed in theoretical space. In cybernetic space we can meet equipment limitations. Here, we meet some discrepancy in considerations on natural, real life and its abstract or artificial equivalent.

## 7 Defining Life

### 7.1 Regulators, Negative Feedbacks and the Homeostat as a Typical Form of Purposeful Information

We can expect, that one of the basic duties of the purposeful information collected in an object is maintaining proper parameters which allow an object to exist. In technology, such mechanisms are called regulators and their construction is based on negative feedbacks. A complete set of regulators is a homeostat, characteristic for living objects. On such basis Bernard Korzeniewski (2001,2005) develops his definition of life. It seems to me that purposeful information not only concerns active regulation but also some other choices of object structure, such as through the use of some substrates which are more available in an environment. However, just the process of collecting and the maintenance of the purposeful information by a *construction* can be described as a negative feedback.

Feedbacks are typical elements of object structure. For the investigation of *structural tendencies* in object changeability, a model of RSN (Gecow 2011,2010a,b,2008a) mentioned above was designed. It describes *complex system*. Feedbacks constituted some of the main problems which the simulation of a complex system must overcome. But these feedbacks are crucial for the complexity needed by a structural tendencies mechanism.

In RSN models, contrary to the *construction*, an object is an active part of the situation $u \in U$. It is a code which transforms the given environment *x* to the subsequent situation containing, by assumption, an object *y* in the form of its properties vector, like in (Gecow 2009b). This *y* is compared to its ideal *y*\* and, based on this comparison, its aptness is calculated. The second part of the resultant situation is an environment which is ignored by the assumption of pass over (see chapter 2.2).

### 7.2 The Deductive Method of Defining Natural Life

Briefly, the described path of designing a life-like phenomenon description has a deductive form based on some axioms. These axioms, our assumptions, are based outside of biology area. The start point is a remark that purposefulness has a right to exist only in life process, and the question outgrowing from this: how may the collecting of purposeful information looks like, if it is not (if it must not to be) an artificial process, *i.e.* if it can be spontaneous.

### 7.3 Dynamics Elements in Information and Coding

Commonly used notions of information, relation and code, do not contain in their nature any connection to time and its direction, but the new notion *natural code* and *purposeful information* exhibit connection to the time direction of transformation by physical laws. This connection, natural in informatics and weakly seen in mathematics, is a basis of the algorithmic information theory (Chaitin 1987, 1979). This theory may be more appropriate to describe my view, but it requires modification in order to 'encode' my view therein. The passage of time is indispensable for causality, for the definition of purpose and process description, and this is my approach.

### 7.4  A Definition of the Natural Life Process?

Is the **spontaneous, long and effective process of collecting purposeful information** the natural life process?

We obtain properties of such process that are very similar to those observed in natural life. I hope, that this is an adequate approach to understanding and defining the natural life process.

An indication of the basic purpose of living objects - '*continuation of existence*' results directly from the assumption of spontaneity of the collecting purposeful information. That is because all: goal, experimenter, information carrier and experimental subject must be the same, single object. An external information carrier can contain different independent information and its variants, but an internal one only one, coherent version. Therefore, if it has already collected some information, then it is a limit of the change size which creates the *natural identity criterion* of the object and allows the object to evolve. For the purpose 'continuation of existence', the *construction* needed for the creation of *purposeful information* occurs to be the Darwinian mechanism. Creation of purposeful information is a getting it from the *natural code*. Therefore, generally, it needs a 'trial and error' method. The measurement of parameters, *e.g.* by a space probe, also is a collection of purposeful information. It is without 'trial and error', but it is only a supplement to the already known physical laws which, used together, can result in a successful landing on an observed planet. The mechanisms of observation and memorizing its effects must have evolved earlier. Now, they are ready to be used and they can be used for artificial collecting, but they cannot be ready at the beginning of the spontaneous process.

Next, *reproduction* is an effect of this special purpose, therefore indirectly of spontaneity. This reproduction (an object itself) cannot be replaced with production (not itself) or an algorithm of coming back to the state before the erroneous hypothesis and elimination. If the experimenter, with his goal and the information carrier can be outside of the experimental system, then the goal can be anything, and the existence of the experimenter and collected purposeful information does not depend on experimental effects and reproduction is not needed.

To treat the above sentence as a definition of life, we should know whether it is not too wider. For this the problem of 'river paradox' (see ch.4.4) should be sufficiently solved.

The question whether a given object is alive or not, seemingly concerns the object, not the process in which it occurs. However, as an effect of reproduction, this process is not a simple process of one object, but a complex process of many partially independent objects and their processes on many levels of their aggregation, disappearance and appearance. Defining an object in such a global process is always a better or worse approximation. First of all, the causes of existence of the given object should be investigated. In many cases, objects exist within the natural process of life, and could not have appeared without it. Like Chaitin (1979), I also 'agree with Orgel (1973) that a definition of 'life' is valid as long as anything that satisfies the definition and is likely to appear in universe under consideration, either is alive or is a by-product of living beings or their activities.' I propose that we go to the next step: it is irrelevant, what we want to call 'alive'; all elements of life process are similarly alive, if they can exist (practically) only as elements of this process.

In this sense a hammer, computer or a bicycle are alive because they all are parts of the natural process of life. However, from the human point of view, subjectively, these artifacts are parts of inanimate matter, and in particular, these objects cannot reproduce. Nevertheless, we have more and more of them in the world around us, which is the result of the properties of these objects and the environment in which they exist. What is the difference between this increase of number and a reproduction? – Only the role of the environment, because we are there in it.

### 7.5 The Different Task of Life Definitions

Most attempts at defining life are going to verify whether a given object is alive or not. I am afraid, that this popular and expected task of life definition is not adequate, generally because it attempts to define an object when it should be a process definition. The proposed life definition does not attempt to realize this task. The main task of this paper is to give some coherent set of notions useful to better understanding what is life and what questions are adequate in this area. In this new perspective many phenomena are looked at in a different way.

### 7.6 Material and Energetic Aspects of Life Specifics

Usually, the physical character of natural life is stressed. It is metabolism for Maynard Smith and Szathmáry (1999) or 'dissipation'. A dissipative system (or dissipative structure) is a thermodynamically open system which operates far from the thermodynamic equilibrium in an environment with which it exchanges energy and matter. As could be seen, for the main properties of life, assumptions of connections to energy, thermodynamic and matter are needless. They may be needed to perform (to find) life process in the real, physical word. This will be the next step of analysis. The distinguish of these two steps leads to two levels of understanding of natural life, the abstract and the physical. A clean, statistical entropy extracted from thermodynamic, which is another view of information and probability, is sufficient for an abstract description of life (its main properties). However, we find that in this 'abstract' case the physical *environment capacity barrier* plays an important role in increasing complexity. Also *cycles* of different *object forms* are an effect of growth after physical reproduction.

### 7.7 Importance of Fuzzy Boundary of Notions

In above attempt to life description we frequently are forced to use simplified unique view like single-valued code and identity or assumption of pass over. In these three cases we clearly know that they are simplifications but problem is that such simplifications remove most important elements of considered phenomena. It is easier to understand that natural identity must and can allow small changes. Code cannot be single-valued for object which must be able to better fulfil goal, and in this way, to collect purposeful information in lot of steps. Possibility of purpose information saving when code and object identity are multi-valued is less intuitive. Purposeful information is defined in one particular code - that one, in which it was tested, therefore constant code is assumed. But changeability of code - environment is one of the most important factors of considered process. It creates *e.g.* covering or strong improvement.

Statistical nature of life phenomenon uses such fuzzy boundary of notions which are clear in other areas, to hide basic qualitative solutions in unexpected cracks in such boundaries. This way the main problems with description and understanding are created.

In descriptions in natural way simplifications appear. They are frequently hard to observe, but they lead to easily observed contradictions. Due to our nature we use particular clear notions, therefore it seems to be impossible to avoid description problems created by fuzzy boundary of real elements of such statistical process. We must remember the above while investigating life.

## 8  The Importance of Differences Between Natural and Artificial Life

The main, and only difference between natural and artificial life is their respective origin. The main characteristics of the process of natural life are derived from an expectation for that process to be spontaneous. These are a goal of '*continuation of existence*', a *natural identity criterion* allowing changeability for constructing hypotheses, eliminating test and reproduction (an object itself). This reproduction cannot be replaced with production (not an object itself) or an algorithm of returning to a state before the erroneous hypothesis and elimination as is done in some simulations of artificial life. Together, these are a Darwinian mechanism of natural selection. In effect 'the **spontaneous, long and effective process of collecting purposeful information**   looks like definition of the natural life process. Along the way, purposeful (biological) and hereditary information is defined, which is the most similar view to that developed by Jablonka et. al.

The process of natural life needs to be an inherent part within a given system, and to be inherently driven by the physical laws of the system. However, from these physical laws only time with causality are needed as the base, and the limited capacity of the environment is used for competition.

The artificial life approach requires a creator which may include all these properties in its result, but it may also define others. If this creator is only a human (or another intelligent creature being an effect of natural life, *e.g.* on other planet) then its artificial life is a part of natural life, which formed its creator. The modelling of artificial life to deeper understanding of natural life must take this under consideration.

## References

Baker, L. R. 2008. The shrinking difference between artifacts and natural objects. *APA Newsletter on Philosophy and Computers* 07(2):2-5.

Chaitin, G. J. 1979. Toward a mathematical definition of "life". In R. D. Levine and M. Tribus, eds. *The Maximum Entropy Formalism*, pages 477–498. MIT Press,.

Chaitin, G. J. 1987. *Algorithmic Information Theory*. Cambridge University Press, Cambridge

Dawkins R. (1976). *The Selfish Gene*. New York City: Oxford University Press. ISBN 0-19-286092-5.

Eigen, M. 1971. Selforganization of matter and the evolution of biological macromolecules, *Naturwissenschaften*, nr 10, pp. 465-523.

Gecow, A., M. Nowostawski, M. Purvis, 2005. Structural tendencies in complex systems development and their implication for software systems. J.UCS, 11,2, 327-356.

Gecow, A. 2008a. Structural Tendencies - effects of adaptive evolution of complex (chaotic) systems. *Int.J Mod.Phys. C* 19, 647-664.

Gecow, A. 2008b. The purposeful information. On the difference between natural and artificial life. *Dialogue & Universalism* 18, 191-206.

Gecow, A. 2009a. Emergence of Growth and Structural Tendencies During Adaptive Evolution of System. in *From System Complexity to Emergent Properties*. M.A. Aziz-Alaoui, Cyrille Bertelle (eds), Springer, Understanding Complex Systems Series, 211-241

Gecow, A. 2009b. The Simple Model of Living Object as an Outside State of Statistical Stable Equilibrium, the Small Change Tendency in Adaptive Evolution.
in *Modelling and Analysis of Complex Interacting Systems*. M.A. Aziz-Alaoui, Cyrille Bertelle (eds), DCDIS-B special issue 515-533

Gecow, A. 2010a. More Than Two Equally Probable Variants of Signal in Kauffman Networks as an Important Overlooked Case, Negative Feedbacks Allow Life in the Chaos. http://arxiv.org/abs/1003.1988

Gecow, A. 2010b.Complexity Threshold for Functioning Directed Networks in Damage Size Distribution http://arxiv.org/abs/1004.3795

Gecow, A. 2011. Emergence of Matured Chaos During Network Growth, Place for Adaptive Evolution and More of Equally Probable Signal Variants as an Alternative to Bias p. In: *Chaotic Systems*, E. Tlelo-Cuautle (ed.), InTech, 280-310. ISBN: 978-953-307-564-8. Available from: www.intechweb.org or http://www.intechopen.com/articles/show/title/emergence-of-matured-chaos-during-network-growth--place-for-adaptive-evolution-and-more-of-equally-probable-signal-variants-as-an-alternative-to-bias-p

Godfrey-Smith, P, K. Sterelny. 2007. Stanford Encyclopedia of Philosophy - Biological Information http://plato.stanford.edu/entries/information-biological/ First published Thu Oct 4, 2007

Griffiths, P., R. Gray. 1994. Developmental Systems and Evolutionary Explanation. *Journal of Philosophy* 91, 277-304.

Griffiths, P. E., R. D. Gray. 2001. Darwinism and Developmental Systems. In *Cycles of Contingency*. S. Oyama, P. E. Griffiths, R. D. Gray (eds),.Cambridge, MA., MIT Press, 195–218.

Jablonka E., M. J. Lamb. 1989 The inheritance of acquired epigenetic variations. *J.Theor.Biol*.139, 69-83.

Jablonka E., M. J. Lamb. 1995. *Epigenetic Inheritance and Evolution: The Lamarckian Dimension*. Oxford University Press.

Jablonka, E. 2002. Information: Its Interpretation, Its Inheritance and Its Sharing. *Philosophy of Science* 69: 578-605.

Jablonka E, Lamb M. J., 2005. *Evolution in four dimensions: genetic, epigenetic, behavioral and symbolic variation in the history of life*. Cambridge, MIT Press.

Jablonka E., 2007. The developmental construction of heredity. *Dev Psychobiol*. 49(8), 808-817.

Jablonka E., Raz G., 2009. Transgenerational Epigenetic Inheritance: prevalence, mechanisms and implications for the study of heredity and evolution. *Quart.Rev.Biol*. 84, 131–176.

Korzeniewski, B. 2001. Cybernetic formulation of the definition of life. *J.Theor.Biol*. 209,275-286

Korzeniewski, B. 2005. Confrontation of the cybernetic definition of living individual with the real word. *Acta Biotheoretica* 53, 1–28 Springer

Küppers, B-O. 1986. *Der Usprung biologischer Information. Zur Naturphilosophie der Lebensentstelung*. R.Piper Gmbh & KG., München.

Lahav, N. 1999. *Biogenesis, Theories of life's origin*. New York-Oxford OUP.

McKay, C. P. 1991. Urey Prize lecture: Planetary evolution and the origin of life. *Icarus*: 92–100.

McKay, C. P. 2004. What Is Life—and How Do We Search for It in Other Worlds? *PLoS Biol* 2(9): e302.

Maynard Smith, J., E. Szathmáry, 1999. *The origin of Life, From the birth of life to the origin of language*. Oxford University Press.

Millikan, R. 1984. *Language, Thought and Other Biological Categories*. Cambridge MA: MIT Press.

Nitecki, M.H. ed. 1988. *Evolutionary Progress*, University of Chicago Press, In this especially: M. H. Nitecki, Discerning the criteria for concepts of progress. pp 3-24. and S. J. Gould, On replacing the idea of progress with an operational notion of directionality. pp 319-338.

Nowostawski M., Gecow A., 2011. *Identity criterion for living objects based on the entanglement measure.* in *Semantic Methods for Knowledge Management and Communication,* Katarzyniak et al. (Eds.) Springer, Studies in Computational Intelligence 381, 159-170

Orgel, L. E. 1973. *The Origins of Life: Molecules and Natural Selection.* Wiley, New York

Pittendright, C. S. 1958. Adaptation, Natural Selection and Behaviour. In *Behaviour and Evolution* ed. A. Roe and G. G. Simpson. New Haven

Rizzotti, M. (red.) 1996. *Defining life: the central problem in theoretical biology*. University of Padova

Shea, N. 2007. Representation in the genome and in other inheritance systems. *Biology and Philosophy* **22,** 313–331

Shearer, W. 1992. *The Atlantic Salmon*. Halstead Press.

Sterelny, K., P. E. Griffiths. 1999. *Sex and Death.* University of Chicago Press.

Taylor, T. J. 1999. *From Artificial Evolution to Artificial Life*. PhD University of Edinburgh http://www.tim-taylor.com/papers/thesis/html/node21.html , ... /node30.html , .../node12.